\documentclass[a4paper,11pt]{article}

\usepackage{tikz}
\usetikzlibrary{backgrounds}
\usepackage{caption}

\usepackage{amsmath,amssymb}    
\usepackage[T1]{fontenc} 
\usepackage[textsize=footnotesize,textwidth=2.5cm]{todonotes}

\usepackage{comment}

\usepackage{color}
\usepackage{graphicx}
\usepackage{subfigure}
\usepackage{cite}               
\usepackage{hyperref}           

\numberwithin{equation}{section}   

\def \be {\begin{equation}}
\def \ee {\end{equation}}
\def \ba {\begin{array}}
\def \ea {\end{array}}
\def \bea{\begin{eqnarray}}
\def \eea{\end{eqnarray}}

\def \d {\delta}
\def \D {\Delta}

\def \m {\mu}
\def \n {\nu}

\def \l {\lambda}

\def \p {\partial}

\def \nn {\nonumber}

\def \mO{\mathcal{O}}

\def \hs {\hspace}

\def \Tr {{\textrm{Tr}}}

\newcommand{\bra}{\left\langle}
\newcommand{\ket}{\right\rangle}

\begin{document}
\title{2d Galilean Field Theories with Anisotropic Scaling}
\author{
Bin Chen$^{1,2,3}$,
Peng-Xiang Hao$^1$
and Zhe-fei Yu$^1$
}
\date{}

\maketitle

\begin{center}
{\it
$^{1}$Department of Physics and State Key Laboratory of Nuclear Physics and Technology,\\Peking University, 5 Yiheyuan Rd, Beijing 100871, P.~R.~China\\
\vspace{2mm}
$^{2}$Collaborative Innovation Center of Quantum Matter, 5 Yiheyuan Rd, Beijing 100871, P.~R.~China\\
$^{3}$Center for High Energy Physics, Peking University, 5 Yiheyuan Rd, Beijing 100871, P.~R.~China}
\vspace{10mm}
\end{center}

\begin{abstract}

In this work, we study two-dimensional Galilean field theories with  global translations and anisotropic scaling symmetries. We show  that such theories have enhanced local symmetries, generated by the infinite dimensional spin-$\ell$ Galilean algebra with possible central extensions, under the assumption that the dilation operator is diagonalizable and has a discrete and non-negative spectrum. We study the Newton-Cartan geometry with anisotropic scaling, on which the field theories could be defined in a covariant way. With the well-defined Newton-Cartan geometry we  establish the state-operator correspondence in anisotropic GCFT, determine the two-point functions of primary operators,   and discuss the modular properties of the torus partition function which allows us to derive Cardy-like formulae.


\end{abstract}

\baselineskip 18pt
\thispagestyle{empty}

\newpage
\tableofcontents
\newpage

\section{Introduction}

In two-dimensional(2D) spacetime, the global symmetry in a quantum field theory  could be enhanced to a local one. The well-known example  studied by J. Polchinski
in \cite{Polchinski:1987dy} shows that a 2D Poincar\'e invariant QFT with scale invariance could be of  conformal invariance, provided that  the theory is unitary and  the dilation spectrum is discrete and non-negative. More recently,  A. Strominger and D. Hofman relaxed the requirement of Lorentz invariance and studied the enhanced symmetries of the theory of chiral scaling\cite{Hofman:2011zj}. They found two kinds of minimal theories. One kind is the two-dimensional conformal field theory (CFT$_2$)\cite{Belavin:1984vu}, while the other kind is called the warped conformal field theory (WCFT). In a warped CFT, the global symmetry is  $SL(2,R)\times U(1)$, and it is enhanced to an infinite-dimensional  group generated by an Virasoro-Kac-Moody algebra. For the study on various aspects of 2D warped CFT, see \cite{Detournay:2012pc,Hofman:2014loa,Castro:2015csg,Castro:2015uaa,Song:2016gtd,Song:2017czq,Jensen:2017tnb,Azeyanagi:2018har,Apolo:2018eky,Chaturvedi:2018uov,Apolo:2018oqv,Song:2019txa}. 

In this paper, we would like to investigate other types of two dimensional field theory with enhanced symmetries. We will focus on the theories whose
 global symmetries include the translations along  two directions, boost symmetry and anisotropic scaling symmetry. If the two directions are recognized as temporal and spatial directions, the anisotropic scaling is of Lifshitz type $x\rightarrow\lambda x,\ t\rightarrow\lambda^z t$. Recall that the scaling behaviour in a warped conformal field theories is chiral
\begin{equation}
x\rightarrow\lambda x,\ \ \ y\rightarrow y,
\end{equation}
while the one in a Galilean conformal field theories (GCFT) is\footnote{In some literature, it sometimes referred the Galilean conformal field theory to be the one with anisotropic scaling $t \to \l^2 t$ and $x_i \to \l x_i$, in particular in higher dimensions\cite{Hagen:1972pd}. With the particle number symmetry, the algebra becomes Schr\"{o}dinger algebra. Without the particle number symmetry, the anisotropic scaling could be $t \to \l^z t$ and $x_i \to \l x_i$ with $z\neq 2$. In this work, we focus on the two-dimensional case, and call the GCFT  the one with $z=1$ and the anisotropic GCFT  the one with $z\neq 1$. }
\begin{equation}
x\rightarrow\lambda x,\ \ \ y\rightarrow \lambda y.
\end{equation}
In Galilean CFT, the boost symmetry is of Galilean type rather than Lorentzian type
\begin{equation}
y\rightarrow y+v x.
\end{equation}
The Galilean CFT can be obtained by taking the non-relativistic limit of the conformal field theory.
Thus the Lorentzian symmetry is broken in Galilean CFT. In this work, we concern the case with more general anisotropic scaling
\begin{equation}
x\rightarrow\lambda^c x,\ \ \ y\rightarrow \lambda^d y,
\end{equation}
and a Galilean boost symmetry. Our consideration is general enough to include the WCFT and GCFT as special cases.

The CFT with anisotropic scaling could be  related to the strong-coupling systems in the condensed matter physics and in some statistical systems\cite{Henkel:1997zz,Henkel:2002vd,Rutkevich:2010xs}. In particular, It is well-known that  for the fermions at unitarity which could be realized experimentally using trapped cold atoms at the Feshbach resonance\cite{Bartenstein:2004zza,Regal:2004zza,Zwierlein:2004zz},  there is Schr\"{o}dinger symmetry, and near the quantum critical points\cite{Sachdev2011} there is Lifshitz-type symmetry.  In order to study these non-relativistic strong coupling systems holographically, people has tried to establish their gravity duals\footnote{ For a nice review and complete references, please see \cite{Hartnoll:2009sz}.} \cite{Son:2005rv, Balasubramanian:2008dm,Kachru:2008yh}. One essential requirement is the geometric realization of the symmetry.

 For a 2D QFT with enhanced symmetry, its role in the holographic duality becomes subtler and more interesting. In this case, the dual gravity must involve 3D gravity. As it is well-known, there is no local dynamical degree of freedom in 3D gravity, but there could be boundary global degrees of freedom. The AdS spacetime is not globally hyperbolic and the boundary conditions at  infinity plays an important role. For AdS$_3$ gravity, under the Brown-Henneaux boundary, the asymptotic symmetry group is generated by two copies of the Virasoro algebra\cite{Brown:1986nw}, leading to the AdS$_3$/CFT$_2$ correspondence. However there exist  other sets of  consistent boundary conditions. In particular, under the Comp\'ere-Song-Strominger boundary conditions, the asymptotic symmetry group is generated by the Virasoro-Kac-Moody U(1) algebra\cite{Compere:2013bya}. Therefore under the CSS boundary conditions, the AdS$_3$ gravity could be dual to a warped conformal field theory. This AdS$_3$/WCFT correspondence has been studied in \cite{Song:2016gtd,Apolo:2018eky,Castro:2017mfj,Chen:2019xpb,Lin:2019dji}. The study of consistent asymptotical boundary conditions and corresponding asymptotic symmetry group have played important roles in setting up other holographic correspondences beyond AdS/CFT, including  chiral gravity\cite{Li:2008dq}, WAdS/WCFT\cite{Anninos:2008fx,Compere:2009zj}, Kerr/CFT\cite{Guica:2008mu}, BMS/GCA\cite{Bagchi:2010eg,Bagchi:2012cy}, BMS/CFT\cite{Barnich:2010eb,deBoer:2003vf,Ball:2019atb} and the non-relativistic limit of the AdS/CFT\cite{Bagchi:2009my}. Recall that both WCFT and GCA are the special cases in our study, it is tempting to guess that the anisotropic GCFT could be the holographic dual of a gravity  theory. In order to investigate this possibility, one needs to study the enhanced symmetry of the field theory and in particular the geometry on which  the theory is defined.

We first study the enhanced symmetries, following the approach developed in \cite{Polchinski:1987dy, Hofman:2011zj}. We find that even with anisotropic scaling and Galilean boost symmetry  there are still infinite conserved charges, equipped with the infinite dimensional spin $\ell=\frac{d}{c}$ Galilean algebra, in the theory. This algebra is different from the chiral part of the $W_\ell$ algebra, even though the weights of the conserved currents are the same. 

 The next question we address is  on what kind of geometry such theories should be defined. Can the local Lorentz symmetry be consistent with the scaling symmetry such that the theories are defined on the pseudo-Riemannian manifold? The answer is generally no. Since the Lorentz boost put the two directions on the equal footing, only the isotropic scaling could be consistent with Lorentz symmetry. Actually as shown in \cite{Sibiryakov:2014qba}, the isotropic scaling may imply the Lorentz  invariance, under the assumption that the propagating speed of signal is finite and  several other assumptions. The existence of isotropic scaling and Lorentz symmetry may lead to 2D CFT defined on the Riemann surfaces. In 2D CFT, the combination of $L_0$ and $\bar{L}_0$ gives the dilation and Lorentz boost generator. On contrast, although 2D GCFT has the isotropic scaling, the propagating speed in it is infinite and the Lorentz invariance is broken as well. For the theories without Lorentz invariance, the geometry cannot be pseudo-Riemannian.

Considering the loss of the local Lorentz symmetry, a natural alternative to pseudo-Riemannian geometry is the Newton-Cartan geometry. In \cite{Hofman:2014loa}, it was noted  that with the global translation and scaling symmetry, the restriction of Lorentz symmetry require the theory to be conformal invariant while the restriction of Galilean symmetry require the theory to be the warped conformal field theories. The warped CFT are defined on the warped geometry, which is a kind of the Newton-Carton geometry with additional scaling structure. For a Galilean invariant field theory\footnote{For various kinds of the Galilean field theories, please see\cite{Duval:2011mi,Martelli:2009uc}. A brand-new application is the discussion on the gravitational waves using the Newton-Carton framework\cite{Duval:1990hj,Zhang:2019gdm}. We thank A. Bagchi and P. A. Horvathy for bringing this point to our attention.}, it could be coupled to a Newton-Cartan geometry in a covariant way\cite{Son:2008ye,Son:2013rqa,Jensen:2014aia,Hartong:2014pma,Banerjee:2014pya,Banerjee:2014nja,Banerjee:2016laq,Duval:2009vt,Duval:1984cj}. For a 2D Galilean conformal field theory, it is expected to couple to a Newton-Cartan geometry with a scaling symmetry, but a detailed study is lacking.
For the Galilean CFT with anisotropic scaling discussed in this paper, we show that it should be defined on a Newton-Cartan geometry with additional scaling structure, similar to the warped geometry discussed in \cite{Hofman:2014loa}. These geometries are actually of  vanishing curvature and non-vanishing torsion.

One advantage of coupling the field theory to geometry is that  the symmetries of the theory become manifest. The theories are defined by requiring the classical action is invariant under certain coordinate reparametrization. For 2D CFT, the Virasoro symmetries is manifest as the worldsheet reparametrization invariance.
For the background Newton-Cartan geometry, the coordinate reparametrization can be absorbed by the local scaling transformation as well as the local Galilean boost. In other words, the theories are defined on the equivalent classes of the Newton-Carton geometry with special scaling structure. The geometries related by local scaling and Galilean boost belong to the same equivalent class\footnote{However, there are potential anomalies in the partition function, since the measure will change under the local transformations. We leave this point to the future work.}.
Having defined these theories, we  find the infinitely many conserved charges by considering the currents coupled to the geometric quantities. These conserved charges are exactly the ones got by using the method in \cite{Hofman:2011zj}.

Furthermore, we study the radial quantization and the state-operator correspondence in the anisotropic GCFT with anisotropic scaling ratio $\ell$ being integer,  analogous to the usual CFT$_2$ case. Remarkably, the primary operators in the theory with $\ell>1$ have different properties. They are not transformed covariantly under the local transformations.  Consequently  the correlation functions become much more complicated than the usual cases.

The Newton-Carton geometry with additional scaling structure has a natural conformal compactification which is of a cylinder  topology  $R\times S^1$. One has a well-defined torus in this case. This allows us to study the modular properties of the theories. We formulate the modular transformation and obtain  the Cardy-like formulae in the general $\ell$ cases. For the warped CFTs with $\ell=0$, and Galilean conformal field theory with $\ell=1$, we get consistent results with the ones in the previous studies.

%


The remaining parts are organized as follows. In Section 2, we generalize the Hofman-Strominger theorem to the anisotropic GCFT. Assuming that the dilation spectrum is discrete and non-negative, the theories coupled to Newton-Carton geometry with global translation and scaling symmetries have infinitely conserved charges. This means the global symmetries are enhanced to local ones. In Section 3, we discuss the properties of the Newton-Cartan geometry with additional scaling structure, on which our field theory could be consistently defined.  It turns out that the geometries should have vanishing curvature but non-vanishing torsion. In Section 4, we give an intrinsic definition of these field theories, from which one can find the allowed local transformations and the corresponding infinitely many conserved charges directly. These discussions match the results in Section 2. In Section 5, we look further into these theories by considering the Hilbert space and the representation of the algebra. The  state-operator correspondence is established. We also discuss the unusual properties of the primary operators for the $\ell>1$ cases. In Section 6, we calculate the two-point functions of the primary operators. A byproduct is the correlation functions of the certain related descendant operators. In Section 7, we define the torus partition function and discuss its  modular properties. We derive a Cardy-like formula which gives estimation of the integral spectrum density at high energy. For a unitary theory, the formula counts the degeneracy of the highly excited states. We conclude and give some discussions in Section 8. 

\section{Enhanced Symmetries}

In this section, we discuss the enhanced symmetries in two dimensional (2D) field theory with boost symmetry and anisotropic scalings, using the method developed in \cite{Polchinski:1987dy} and \cite{Hofman:2011zj}. Usually for a theory with   global symmetries, we can  defines the corresponding conserved Noether currents and their conserved charges. However there could be ambiguities in defining the currents. In 2D quantum field theory with scaling symmetry and boost symmetry, under the assumption that there exist  a complete basis of local operators as the eigenvectors of the dilation operator with a discrete spectrum, the conserved currents can be organized in a form such that they have the canonical commutation relations with the generators. But the currents can be shifted by  certain local operators without changing the commutation and conservation relations. Analyzing the behavior of the local operators leads to special relations of the currents, which in turn tell us that there may be infinite conserved charges.

\subsection{Global symmetries}

The global symmetries of 2D QFT we consider in this work include the translations along two directions $x$ and $y$
\begin{equation}
x\rightarrow x'=x+\delta x,
\hs{3ex}
y\rightarrow y'=y+\delta y,
\end{equation}
the dilations
\begin{equation}
x\rightarrow x'=\lambda ^c x,\hs{3ex}y\rightarrow y'=\lambda ^d y,
\end{equation}
where $c,d$ are non-negative.
And the Galilean boost symmetry behaves as,
\begin{equation}
y\rightarrow y'=y+v x.
\end{equation}
It is worth noting that the dilation scales two directions at the same time, but could be with different weights $c$ and $d$.  We use a slightly different notation from the one in \cite{Hofman:2014loa}. The generators of the above symmetry transformations are denoted as $H, \bar{H}, D$ and $B$ respectively. They annihilate the vacuum, and satisfy the  commutation relations
\bea
[H,\bar{H}]=0,&[D,H]=-c H,&[D,\bar{H}]=-d \bar{H},\eea
\bea
[B,H] =-\bar{H},&[B,\bar{H}]=0,&[B,D]=(d-c)B.
\eea

We assume that the dilation operator has a discrete spectrum and  the theory has a
 complete basis of local operators which obey
\bea
[H,\mO]=\partial_x \mO,&&[\bar{H},\mO]=\partial_y \mO,\nn \eea
\bea
[D,\mO]=c x \partial_x \mO+d y \partial_y \mO+\Delta_\mO \mO,
\eea
where $\D_\mO$ is the non-negative scaling dimension of the operator $\mO$. The global symmetries can restrict the  two-point function of $\mO_1,\mO_2$ to be either of the form
\begin{equation}
\bra \mO_1(x_1,y_1) \mO_2(x_2,y_2)\ket=x_{12}^{-c(\Delta_1+\Delta_2)} f(\frac{y_{12}^c}{x_{12}^d}),
\end{equation}
or of the form
\begin{equation}
\bra \mO_1(x_1,y_1) \mO_2(x_2,y_2)\ket=y_{12}^{-d(\Delta_1+\Delta_2)} f(\frac{x_{12}^d}{y_{12}^c})
\end{equation}
where $x_{12}=x_1-x_2,\ y_{12}=y_1-y_2$  and $f$ is a priori unknown function.
Moreover in the case that the operators are invariant under the Galilean boost
\begin{equation}
[B,\mO(x,y)]=x\partial_y \mO(x,y),
\end{equation}
the two-point function of $\mO$ does not depend on $y_{12}$
\begin{equation}
\bra \mO(x_1,y_1) \mO(x_2,y_2)\ket=N_{\mO}x_{12}^{-2c\Delta_{\mO}},
\end{equation}
where $N_{\mO}$ is the normalization constant. Here for simplicity, we take $\mO_1=\mO_2$.

The generators above are related to the conserved Noether current by
\begin{equation}
Q=\int \star J,
\end{equation}
where
\begin{equation}
\star =H_{\mu\nu}
\end{equation}
 serves as the volume in the Newton-Cartan geometry which will be studied in the next section, $J$ is the conserved current satisfying
\begin{equation}
\nabla_{\mu}J^{\mu}=0.
\end{equation}
In  flat Newton-Cartan geometry,
\begin{equation}
Q=\int J_x dx+\int J_y dy, \hs{3ex}\mbox{with}\hs{2ex}\partial_y J_x+\partial_x J_y=0.
\end{equation}
The integral contour is the slice where we quantize the theory and define the Hilbert space. 

Corresponding to the generators $H, \bar{H}, D$ and  $B$, the currents are denoted as $h_\m,\bar{h}_\m,d_\m,b_\m$. The canonical commutation relations of the currents and the charges are
\be
[H,h_x]=\partial_x h_x,\hs{2ex} [H,h_y]=\partial_x h_y,\hs{2ex}[H,\bar{h}_x]=\partial_x\bar{h}_x,\hs{2ex}[H,\bar{h}_y]=\partial_x\bar{h}_y,
\ee
\be
[H,d_x]=\partial_x d_x+c h_x,\hs{2ex}[H,d_y]=\partial_x d_y+c h_y,\ee
\be
[H,b_x]=\partial_x b_x+\bar{h}_x,\hs{2ex}[H,b_y]=\partial_x b_y+\bar{h}_y,
\ee
\begin{equation}
[\bar{H},h_x]=\partial_y h_x,\hs{2ex}[\bar{H},h_y]=\partial_y h_y,\hs{2ex}[\bar{H},\bar{h}_x]=\partial_y\bar{h}_x,\hs{2ex}[\bar{H},\bar{h}_y]=\partial_y\bar{h}_y
\end{equation}
\begin{equation}
[\bar{H},d_x]=\partial_y d_x+d h_x,\hs{2ex}[\bar{H},d_y]=\partial_y d_y+d h_y,\ee
\be
[\bar{H},b_x]=\partial_y b_x,\hs{2ex}[\bar{H},b_y]=\partial_yb_y,
\end{equation}
\begin{equation}
[D,h_x]=(c x\partial_x+d y\partial_y)h_x+2c h_x,\hs{2ex}[D,h_y]=(c x\partial_x+d y\partial_y)h_y+(c+d) h_y,
\end{equation}
\begin{equation}
[D,\bar{h}_x]=(c x\partial_x+d y\partial_y)\bar{h}_x+(c+d) \bar{h}_x,\hs{2ex}[D,\bar{h}_y]=(c x\partial_x+d y\partial_y)\bar{h}_y+2d \bar{h}_y,
\end{equation}
\begin{equation}
[D,d_x]=(c x\partial_x+d y\partial_y)d_x+c d_x,\hs{2ex}[D,d_y]=(c x\partial_x+d y\partial_y)d_y+d d_y,
\end{equation}
\begin{equation}
[D,b_x]=(c x\partial_x+d y\partial_y)b_x+d b_x,\hs{2ex}[D,b_y]=(c x\partial_x+d y\partial_y)b_y+(2d-c) b_y,
\end{equation}
\begin{equation}
[B,h_x]=x\partial_yh_x- h_x,\hs{2ex}[B,h_y]=x\partial_yh_y,\ee
\be
[B,\bar{h}_x]=x\partial_y\bar{h}_x,\hs{2ex}[B,\bar{h}_y]=x\partial_y\bar{h}_y+\bar{h}_y,
\end{equation}
\begin{equation}
[B,d_x]=x\partial_yd_x- d_x,\hs{2ex}[B,d_y]=x\partial_yd_y,\ee
\be
[B,b_x]=x\partial_yb_x,\hs{2ex}[B,b_y]=x\partial_yb_y+b_y.
\end{equation}
We choose the above commutation relations by the following two requirements. One is that the differential operators must act on the field properly, while the other is that we must recover the commutators of the generators.

It is remarkable that there are ambiguities in defining the Noether currents. One can shift the currents by some local operators to get the same commutation relations of the generators and still have the conservation laws. One may organize the currents with respect to the  canonical commutation relations to define the local operators. The above canonical commutation relations imply that the dilation and boost currents can be expressed by the translation current up to some local operators
\begin{equation}
d_x=c xh_x+d y\bar{h}_x+s_x,\hs{2ex}d_y=c xh_y+d y\bar{h}_y+s_y
\end{equation}
\begin{equation}
b_x=x\bar{h}_x+w_x,\hs{2ex}b_y=x\bar{h}_y+w_y,
\end{equation}
where $s_{x},s_y$ and $w_{x},w_y$ are local operators. In the following we will study the shifts of the currents that do not change the canonical commutation relations.

\subsection{Enhanced symmetries}

Let us first study the boost symmetry and the boost current. The boost currents are related to the translation currents
\begin{equation}
b_x=x\bar{h}_x+w_x,\hs{2ex}b_y=x\bar{h}_y+w_y.
\end{equation}
The conservation law reads,
\begin{equation}
\partial_y b_x+\partial_x b_y=0,\hs{2ex}\partial_x\bar{h}_y+\partial_y\bar{h}_x=0
\end{equation}
which allows us to write the current $\bar{h}_y$ as,
\begin{equation}
\bar{h}_y=-\partial_x w_y-\partial_y w_x.
\end{equation}
From the commutation relations, we learn that $w_x$ is invariant under the Galilean boost. From the discussion on the two-point functions, we find
\begin{equation}
\partial_{y_1}\bra w_x(x_1,y_1) w_x(x_2,y_2)\ket =0,\hs{2ex}\bra \partial_y w_x \partial_y w_x\ket=0.
\end{equation}
From our assumptions that the spectrum of the dilation operator is discrete and non-negative, the following equation is valid as an operator equation
\begin{equation}
\partial_y w_x=0.
\end{equation}
We can shift the currents without changing the canonical commutation relations
\begin{equation}
\bar{h}_y\rightarrow \bar{h}_y+\partial_x w_y,\hs{2ex}\bar{h}_x\rightarrow \bar{h}_x-\partial_y w_y.
\end{equation}
The $\bar{h}_x$ component must be changed at the same time to keep the conservation law intact. The similar shifts also happen in the  currents $b_\m$.  Under the above shift, we can set
\begin{equation}
\bar{h}_y=0, \label{hbary}
\end{equation}
such that
\be
\partial_y\bar{h}_x=0
\ee
which implies that $\bar{h}_x$ is a function of $x$
\begin{equation}
\bar{h}_x=\bar{h}_x(x).
\end{equation}
This leads to the existence of an infinite set of conserved charges,
\begin{equation}
M_{\epsilon}=\int \epsilon(x)\bar{h}_x(x) dx, \label{Mcharges}
\end{equation}
where $\epsilon(x)$ is an arbitrary smooth function  $x$. It is easy to see that  $M_1$ with $\epsilon=1$ actually generates the translation along $y$ direction, while $M_x$  with $\epsilon(x)=x$ is the boost generator. This is consistent with the discussion in the warped CFT literatures\cite{Hofman:2011zj,Detournay:2012pc}. We should emphasize here that this infinite set of conserved charges are common in the 2d local Galilean field theories. 

Next let us turn to the dilation current. Depending on the weight $c$,  we will consider $c=0$ and $c\neq 0$  separately.

\subsubsection{Special case: $c=0$}

In this case, we have
\begin{equation}
d_x=d y\bar{h}_x+s_x,\hs{2ex} d_y=dy\bar{h}_y+s_y.
\end{equation}
The equations above can be taken as the defining relations of new local operators $s_x$ and $s_y$, taking into account of the fact that
\begin{equation}
\bar{h}_x=\bar{h}_x(x),\hs{2ex}\bar{h}_y=0.
\end{equation}
The canonical commutation relations are still valid, as well as the conservation laws. Considering the conservation laws of $d_\m$ and $h_\m$ \begin{equation}
\partial_y d_x+\partial_x d_y=0,\hs{2ex}\partial_y h_x+\partial_x h_y=0
\end{equation}
we have
\begin{equation}
d\bar{h}_x=-\partial_y s_x-\partial_x s_y.
\end{equation}
Now $s_x$ is an  operator of weight zero under the dilation. The two-point function is
\begin{equation}
\bra s_x s_x \ket=\mbox{constant},
\end{equation}
which implies that
\begin{equation}
\partial_y s_x=0
\end{equation}
is valid as an operator equation. We arrive at
\begin{equation}
\bar{h}_x(x)=-\partial_x s_y.
\end{equation}
Note that $s_y$ is an operator of  weight $d$ under the dilation along $y$ direction such that
\begin{equation}
\partial_y\bra s_y s_y \ket=f(x)\partial_y y^{-2d}\neq 0.
\end{equation}
But $s_y$ is invariant under the Galilean boost as well, which means that the above relation should be vanishing. The only way to be self-consistent is to set $s_y=0$ and therefore $\bar{h}_x=0$. This implies that  a 2D theory with $c=0$ and
 the symmetries
\begin{equation}
y\rightarrow y+v x,\hs{2ex}y\rightarrow \lambda y,\hs{2ex}y\rightarrow y+\delta y,
\end{equation}
is inconsistent and  does not exist.

\subsubsection{Other cases: $c\neq 0$}

Next we  turn to the $c\neq 0$ cases, in which we can normalize the dilation so that $c=1$. However, we keep $c$ unfixed in the following discussion in this section. One should note that  the final results cannot be symmetric in $c$ and $d$, since the boost symmetry tells the difference between $x$ and $y$ directions.

We start from  the dilation currents $d_\m$
\begin{equation}
d_x=c xh_x+d y\bar{h}_x+s_x,\hs{2ex}d_y=c xh_y+d y\bar{h}_y+s_y.
\end{equation}
The conservation law of $d_\m$ leads to the relation
\begin{equation}
c h_y+d\bar{h}_x=-\partial_y s_x-\partial_x s_y.
\end{equation}
Moreover we have
\begin{equation}
\bar{h}_y=0.
\end{equation}
We can shift the current $h_\m$ as follows
\bea
h_y&\rightarrow &h'_y=h_y +\frac{1}{c} (\partial_y s_x+\partial_x s_y),\nn\\
h_x &\rightarrow & h'_x=h_x -\frac{1}{c} (\partial_x s_x+\partial_y s_y).
\eea
This will not change the commutation relations and the conservation laws.

Considering the boost behaviour of $s_x,s_y$, after the shift we may have
\begin{equation}
c h_y+d\bar{h}_x=0. \label{hy}
\end{equation}
We can define a set of charges,
\begin{equation}
L_\epsilon=\int \{c\epsilon(x) h_x(x,y)+d\epsilon'(x)y\bar{h}_x(x)\}dx+\int\{c\epsilon(x)h_y(x) \}dy, \label{Qcharges}
\end{equation}
where $\epsilon(x)$ is arbitrary smooth function on $x$ and $\epsilon'(x)=\p_x \epsilon$. $h_y(x)$ depends only on $x$, since its boost charge vanishes. We denote
\begin{equation}
  q_x=c\epsilon(x) h_x(x,y)+d\epsilon'(x)y\bar{h}_x(x), \hs{2ex}q_y=c\epsilon(x)h_y(x).
\end{equation}
One can check that the charges $L_\epsilon$ are indeed conserved
\begin{equation}
\partial_y q_x+\partial_x q_y=0,
\end{equation}
provided that
\begin{equation}
\partial_yh_x+\partial_xh_y=0.
\end{equation}
Note that when $\epsilon=1$,
\begin{equation}
L_1=\int h_xdx+\int h_ydy
\end{equation}
generates the translation in $x$ direction, while when $\epsilon=x$
\begin{equation}
L_x=\int \{cxh_x(x,y)+dy\bar{h}_x(x)\}dx+\int\{cxh_y(x) \}dy
\end{equation}
 generates the anisotropic scaling symmetry.

 In the case that $d=0$, from \eqref{hy} we have
 \begin{equation}
 h_y=0.
 \end{equation}
And considering the conservation law, we find that $h_x$ depends only on $x$. This is exactly the case for the warped CFTs discussed in \cite{Hofman:2011zj,Hofman:2014loa}.

\subsection{Algebra of enhanced symmetries}
After some calculations, we arrive at the algebra,
\begin{equation}
[L_\epsilon,L_{\tilde{\epsilon}}]=L_{c \epsilon'\tilde{\epsilon}-c \tilde{\epsilon}'\epsilon}+\cdots,
\end{equation}
\begin{equation}
[L_\epsilon,M_{\tilde{\epsilon}}]=M_{d\epsilon'\tilde{\epsilon}-c \tilde{\epsilon}'\epsilon}+\cdots,
\end{equation}
\begin{equation}
[M_\epsilon,M_{\tilde{\epsilon}}]=\cdots,
\end{equation}
where $\epsilon$ and $\tilde{\epsilon}$ are arbitrarily smooth functions of $x$ and the
 ellipsis denotes potential central extension terms allowed by the Jacobi identity.\\
The algebra of the plane modes without central extension is
  \begin{eqnarray}\label{algebra}
\nonumber \left[l_n,l_m\right]&=& c(n-m)l_{n+m},\\
\nonumber\left[l_n,m_m\right]&=& (dn-c m)m_{n+m} ,\\
\left[m_n,m_m\right] &=&0.
\end{eqnarray}
This is the infinite dimensional spin-$\ell$ Galilean algebra, with $\ell=\frac{d}{c}$\cite{Henkel:1997zz}.

The central extension is constrained by the Jacobi identity\cite{Hosseiny:2014dxa}. There are various kinds of extensions, which we list here in order.
\begin{itemize}
\item $T$-extension is always allowable:
  \begin{equation}
 \left[L_n,L_m\right]= (n-m)L_{n+m}+\frac{c_T}{12}n(n^2-1)\delta_{n+m,0}.
  \end{equation}
  This gives the Virasoro algebra.
    \item $B$-extension is only allowable for $\ell=1$:
  \begin{equation}
 \left[L_n,M_m\right]= (n-m)M_{n+m}+\frac{c_B}{12}n(n^2-1)\delta_{n+m,0}.
  \end{equation}
 This gives the Galilean conformal algebra (GCA). The field theories equipped with GCA have been discussed in \cite{Bagchi:2009ca,Bagchi:2009pe,Bagchi:2016geg,Bagchi:2017cpu}.
    \item $M$-extension is only allowable for $d=0$, the infinite dimensional spin-$0$ Galilean algebra 
  \begin{equation}
\left[M_n,M_m\right]=c_Mn\delta_{n+m,0}.
  \end{equation}
  This is actually the algebra for the warped CFT, with $c_M$ being the Kac-Moody level.
  \item Infinite $M$-extensions, in which  there are infinite $c_M$ charges
  \begin{equation}
[M_n,M_m]=(n-m){(c_M)}_{n+m},\hs{3ex}
[L_n,{(c_M)}_m]=-m{(c_M)}_{n+m}.
  \end{equation}
  The familiar case is the Schr\"{o}dinger-Virasoro algebra, in which $\ell=1/2$. Note that for arbitrary spin $\ell$, there could be similar algebraic structure.
\end{itemize}

\section{Geometry}

In this section, we discuss the underlying geometry on which the theories with anisotropic scaling and boost symmetries can be defined.
Recall that  a 2D $CFT$ in the Euclidean signature is defined on a two-dimensional Riemann surface, which has the translation symmetries, rotation symmetry and a  scaling symmetry.
More importantly the classical action is invariant under the (anti-)holomorphic transformations
\begin{equation}
z\rightarrow f(z),\ \ \ \bar{z}\rightarrow f(\bar{z}),
\end{equation}
but the partition function and correlation functions may suffer from potential quantum anomaly due to the change of the measure under the transformations.

For the Galilean field theories, one needs to introduce the Newton-Cartan structure into the two-dimensional geometry  to make the Galilean symmetries manifest. Furthermore, a special scaling structure is needed to define the dynamical variable, the affine connection. For the warped CFTs, the underlying Newton-Cartan geometry has been studied in \cite{Hofman:2014loa}. For the case at hand, we need to introduce a Newton-Cartan geometry with a different scaling structure however. 

\subsection{Flat Geometry}

We start with the geometry similar to the flat Euclidean geometry. Such geometry admits the following symmetries
\begin{equation}
H:x\rightarrow x'=x+\delta x,
\end{equation}
\begin{equation}
\bar{H}:y\rightarrow y'=y+\delta y,
\end{equation}
\begin{equation}
B:y\rightarrow y'=y+v x.
\end{equation}
Note that for different scalings $c,d$ , the flat geometries are the same.

The invariant vector and one-form are respectively
\begin{equation}
\bar{q}^a=\left(
\begin{aligned}
& 0\\
& 1
\end{aligned}
\right )\ ,\ \ \ \ q_a=(0\ \ \ 1),\hs{3ex}a=1,2.
\end{equation}
Similarly, there is a metric
\begin{equation}
g_{ab}=q_aq_b=\left(
\begin{aligned}
1\ \ & 0\\
0\ \ & 0
\end{aligned}
\right )
\end{equation}
which is flat and invariant under boost transformation
\begin{equation}
g=BgB^{-1}.
\end{equation}
The metric is degenerate, and it is orthogonal to the invariant one-form. It has one positive eigenvalue and one vanishing eigenvalue.
Besides, there is an antisymmetric tensor $h_{ab}$ to lower the index
\begin{equation}
q_a=h_{ab}\bar{q}^b.
\end{equation}
 It is invariant under the boost transformation as well. It is invertible with $h^{ab}h_{bc}=\d^a_c$, and its inverse helps us to raise the index
 \be
 \bar{q}^a=h^{ab}q_b.
 \ee
With $h^{ab}$, we can obtain the upper index  metric
\be
\bar{g}^{ab}=\bar{q}^a\bar{q}^b=h^{ac}h^{bd}q_c q_d=h^{ac}h^{bd}g_{cd}.
\ee

\subsection{Curved Geometry}

In the previous subsection, the vector space and the dual 1-form space are introduced to define the geometry. The antisymmetric tensor $h_{ab}$ maps the vectors to one-forms, and the metric $g_{ab}$ defines the inner product of the vectors. This is in contrast with  the usual Riemannian geometry, in which the metric serves also as a tool to map the vectors to the one-forms.

The curved geometry is defined by `gluing  flat geometry', in the sense that the tangent space is flat  with the map determined by the zweibein. One needs to  define the connection properly. The zweibein is required to map the space-time vector to the tangent vector,
\begin{equation}
e^a_{\mu}:v^{\mu}\rightarrow \bar{v}^a
\end{equation}
The covariant derivative is
\begin{equation}
D=\partial+\omega+\Gamma
\end{equation}
where $\omega$ is the spin connection to connect the points in the tangent space, while $\Gamma$ is the affine connection to connect the points in the base manifold. In the usual case,
 the affine connection is determined uniquely by requiring the metric to be compatible and torsion free, with zweibein postulate.
  In the Galilean case, the torsion free condition cannot determine the spin connection uniquely, and  other conditions should be imposed to get the unique spin connection and then the affine connection by zweibein postulate\cite{Jensen:2014aia,Bergshoeff:2014uea}.
  From the zweibein postulate
\begin{equation}
D_{\mu}e^a_{\nu}=0,
\end{equation}
and the invertibility of $e^a_\m$,
one may get
the affine connection
\begin{equation}
\Gamma^\rho_{\mu\nu}=e^\rho_a\partial_\mu e^a_\nu+e^\rho_a e^b_\nu \omega^a_{~b\mu}
\end{equation}
where
\begin{equation}
\omega^a_{~b\mu}=\bar{q}^aq_b\omega_{\mu}.
\end{equation}
The torsion and curvature two-forms are respectively
\begin{equation}
T^a=de^a+\omega^a_{~b} \wedge e^b, \hs{3ex}
R^a_{~b}=d\omega^a_{~b}.
\end{equation}
The metric compatibility requires,
\begin{equation}
D_\mu \bar{q}^a=D_\mu q_a=0.
\end{equation}
Instead of torsion free condition, the condition proposed here is that the geometry is compatible with the scaling symmetry, i.e. the scaling structure is a covariant constant
\begin{equation}
D_\mu J^a_b=0.
\end{equation}
The scaling structure is defined to select the scaling weights of  vectors and 1-forms
\begin{equation}
J^a_b\bar{q}^b=-d\bar{q}^a, \hs{3ex}
J^a_bq^b=-c q^a.
\end{equation}
 Under the scaling ,
\begin{equation}
x\rightarrow \lambda^c x,\ \ \ y\rightarrow \lambda^dy,
\end{equation}
the infinitesimal transformation is
\begin{equation}
\Lambda^a_b=\delta^a_b+\lambda J^a_b.
\end{equation}
The scaling structure is expressed covariantly as,
\begin{equation}
J^a_b=-c(q^cq_c)^{-1}q^aq_b-d(\bar{q}^c\bar{q}_c)^{-1}\bar{q}^a\bar{q}_b
\end{equation}
by requiring that
\begin{equation}
\bar{q}_aq^a=0.
\end{equation}
Note again that $\bar{q}^a$ and $q_a$ are boost invariant vector and one-form, and then the vector $q^a$ and one-form $\bar{q}_a$ are defined by the scaling structure in turn.
Now the condition  that the scaling structure is covariant constant implies
\begin{equation}
q_a\partial_\mu q^a=0,\ \ \ \bar{q}^a\partial_\mu \bar{q}_a=0,
\end{equation}
which means that
\begin{equation}
q_aq^a=\mbox{const.}, \ \  \ \bar{q}_a\bar{q}^a=\mbox{const.}
\end{equation}
As at different points, the normalization should be the same, one can choose the constants to be unit.

The scaling structure is covariantly constant also implies the spin connection can be expressed as
\begin{equation}
\omega_\mu=-\frac{1}{c+d}(c\bar{q}_a\partial_\mu q^a+dq^b\partial_\mu \bar{q}_b).
\end{equation}
One should also impose that
\begin{equation}
D_\mu q^a=0.
\end{equation}
This means the weights of vectors do not change when being parallel transported. This  condition implies
\begin{equation}
\bar{q}_a\partial_\mu q^a=q^a\partial_\mu \bar{q}_a.
\end{equation}
Then one reaches the conclusion that in the $(x,y)$ coordinates in the tangent space
\begin{equation}
\omega_\mu=0,
\end{equation}
and in turn
\begin{equation}
R=0.
\end{equation}
However, the affine connection and the torsion
\begin{equation}
\Gamma^\rho_{\mu\nu}=e^\rho_a\partial_\mu e^a_\nu,\ \ \ T^a=de^a
\end{equation}
are now not vanishing.
This is the same as the warped geometry of warped CFTs. 

\subsection{Affine connection}

In this subsection, we discuss the various constraints to determine the affine connection without the help of the zweibein.
The starting point is the Newton-Cartan geometry $(M,A_{\mu},G^{\mu\nu})$.
$A_\mu$ is a temporal one-form which defines the local time direction, while $G^{\mu\nu}$ is the inverse metric on the spatial slice.
One may define
\begin{equation}
G^{\mu\nu}=\bar{A}^\mu\bar{A}^\nu,\ \ G_{\mu\nu}=A_\mu A_\nu,
\end{equation}
and the antisymmetric tensor
\be
H_{\m\n}=e^a_\m e^b_\n h_{ab}=e^a_{[\m} e^b_{\n]} h_{ab}=A_{[\m}\bar{A}_{\n]}.
\ee
The velocity field $A^{\mu}$ is defined by
\be
A_\mu\bar{A}^\mu=0, \hs{2ex}\bar{A}^\m A_\m=0,
\ee
where $\bar{A}_\mu$ is the dual one-form of $A^\mu$
\begin{equation}
\bar{A}_\nu=H_{\mu\nu}A^\mu.
\end{equation}
The vectors and one-forms are related to the zweibein in the last subsection by
\begin{equation}
\hat A=\hat e \cdot \hat q.
\end{equation}
In components, we have
\begin{equation}
\bar{A}^\mu=e_a^\mu\bar{q}^a,\ \ \ A_\mu=e^a_\mu q_a,
\end{equation}
\begin{equation}
\bar{A}_\mu=e^a_\mu\bar{q}_a,\ \ \ A^\mu=e_a^\mu q^a.
\end{equation}


The question is what conditions should be imposed to determine the geometry completely. In the following, we review the fact that metric compatibility and the torsion free condition cannot determine the affine connection uniquely.
The covariant derivative acts on the tensor as
\begin{equation}
D_\mu V^\nu_\rho=\partial_\mu V^\nu_\rho+\Gamma^\nu_{\sigma\mu}V^\sigma_\rho-\Gamma^\sigma_{\rho\mu}V^\nu_\sigma.
\end{equation}
The torsion is
\begin{equation}
T^\mu_{\nu\rho}=\Gamma^\mu_{\nu\rho}-\Gamma^{\mu}_{\rho\nu},
\end{equation}
and the curvature is defined as usual.
Then the constancy condition of $A_\mu$ implies
\begin{equation}
D_\mu A_\nu=\partial_\mu A_\nu-\Gamma^\rho_{\mu\nu}A_\rho=0
\end{equation}
which gives constraints on the temporal affine connection
\begin{equation}
\Gamma^\rho_{\mu\nu}A_\rho=\partial_\mu A_\nu.
\end{equation}
Along with the torsion free condition
\begin{equation}
T^\mu_{\nu\rho}=\Gamma^\mu_{\nu\rho}-\Gamma^{\mu}_{\rho\nu}=0,
\end{equation}
one gets the point that the temporal one-form is closed
\begin{equation}
\partial_\mu A_\nu-\partial_\nu A_\mu=0.
\end{equation}
Considering the constancy of $\bar{A}^\mu$, one finds the affine connection
\begin{equation}
\Gamma^\mu_{\nu\rho}=\bar{A}^\mu\bar{A}^\sigma(\bar{A_{(\rho}}\partial_{\nu)}\bar{A}_\sigma-\bar{A}_{(\nu|}\partial_\sigma\bar{A}_{|\rho)})+\bar{A}^\mu\partial_{(\nu}\bar{A}_{\rho)}+\bar{A}^\mu\bar{A}^\sigma A_{(\nu}F_{\rho)\sigma},
\end{equation}
where $F_{\mu\nu}$ is an arbitrary anti-symmetric tensor.
Moreover we impose the condition that the scaling structure is covariant constant, which implies that the parallel transport keeps the scaling weight of the vectors invariant. This fact implies that
\begin{equation}
D_\mu A^\nu=0,
\end{equation}
and then
\begin{equation}
F_{\mu\nu}=0.
\end{equation}
The requirement that the scaling structure is covariantly constant implies also
\begin{equation}
\bar{A}^\mu\bar{A}_\mu=\mbox{const}.
\end{equation}
This in turn determines $A^\mu$ and the affine connection
\begin{equation}
\Gamma^\rho_{\mu\nu}=0.
\end{equation}
Actually the conditions are too strong to allow interesting geometry.

To get the non-vanishing affine connection, one may relax the torsion free condition. The only constraints we impose are the metricity and the condition that the scaling structure is covariantly constant. Then the affine connection reads
\begin{equation}
\Gamma^\rho_{\mu\nu}=A^{\rho}\partial_{\mu}A_\nu+\bar{A}^{\rho}\partial_{\mu}\bar{A}_\nu.
\end{equation}
In this case, the curvature is vanishing, but the torsion tensor is not
\begin{equation}
R=0,\ \ T_{\mu\nu}^\rho=A^\rho\partial_{[\mu}A_{\nu]}+\bar{A}^\rho\partial_{[\mu}\bar{A}_{\nu]}.
\end{equation}
 This is the case we focus on in this paper. Note that if one does not require the covariantly constant scaling structure, there are remaining ambiguities, the so-called Milne boost, in defining the velocity vector.

One may impose another set of consistent constraints, including the metric compatibility, torsion free and
\begin{equation}
R^{(\mu\nu)}_{[\rho\sigma]}=0.
\end{equation}
These conditions
 imply
\begin{equation}
dF=0,\ \ \ F=dQ.
\end{equation}
$F$ is closed and can be expressed as an exterior derivative of a local $U(1)$ connection coupled to the particle number current. This is the so-called the geometry with Newtonian connection. The field theories defined on such geometries have non-vanishing central terms which are the particle numbers or the mass extensions.

\section{Defining  Field Theories}

In this section, we discuss what kinds of field theories could be coupled to the geometry discussed above in a covariant way, and check that there are indeed infinitely many conserved charges in these theories. As the case $c=0$  is  trivial, here we focus on the case $c\neq 0$. To simplify the notation, we use the freedom in the overall re-scaling to set $c=1$.

The geometry is defined by $A_\mu,\bar{A}_\mu,A^\mu,\bar{A}^\mu$, satisfying
\begin{equation}
A_\mu A^\mu=1,\ \bar{A}_\mu\bar{A}^\mu=1,\ A_\mu\bar{A}^\mu=0,\ \bar{A}_\mu A^\mu=0.
\end{equation}
In the discussion below, the canonical one-forms are chosen to be
\begin{equation}
A=dx,\ \ \bar{A}=dy.
\end{equation}
Under the scaling
\begin{equation}
x\rightarrow \lambda x,\ \ y\rightarrow \lambda^dy,
\end{equation}
the vector field and 1-form field transform as
\begin{equation}
A_\mu\rightarrow \lambda A_\mu,\ \ \bar{A}_\mu\rightarrow \lambda^d\bar{A}_\mu,
\end{equation}
\begin{equation}
A^\mu\rightarrow \lambda^{-1} A^\mu,\ \ \bar{A}^\mu\rightarrow \lambda^{-d} \bar{A}^\mu.
\end{equation}
Under the boost
\begin{equation}
y\rightarrow y+v x,
\end{equation}
there is
\begin{equation}
A_\mu\rightarrow A_\mu,\ \ \bar{A}_\mu\rightarrow\bar{A}_\mu+vA_\mu,
\end{equation}
\begin{equation}
A_\mu\rightarrow A_\mu-v \bar{A}^\mu,\ \ \bar{A}^\mu\rightarrow\bar{A}^\mu.
\end{equation}

Now we want to find the diffeomorphism of the geometry by considering an infinitesimal coordinate transformation,
\begin{equation}
x\rightarrow x+\epsilon(x,y),\ \ \ y\rightarrow y+\xi(x,y).
\end{equation}
The infinitesimal variations are,
\bea\label{if1}
\delta dx&=&\partial_x \epsilon dx+\partial_y \epsilon dy,\nn\\
\delta dy&=&\partial_x \xi dx+\partial_y \xi dy.
\eea
This should be the same as the one arisen from the Galilean boost and anisotropic transformations locally,
\bea\label{if2}
\delta dx&=&\lambda dx,\nn\\
\delta dy&=&d\lambda dy+v dx.
\eea
Comparing \eqref{if1} with \eqref{if2}, we get the constraints on the transformations,
\begin{equation}
d\partial_x \epsilon(x,y)=\partial_y \xi(x,y),
\end{equation}
\begin{equation}
\partial_y\epsilon(x,y)=0.
\end{equation}
The allowed infinitesimal transformations are
\begin{equation}
x\rightarrow x+\epsilon(x),\ \ \ y\rightarrow (1+d\epsilon'(x))y, \label{epsilon1}
\end{equation}
\begin{equation}
x\rightarrow x,\ \ \ y\rightarrow y+\xi(x), \label{epsilon2}
\end{equation}
It turns out the allowed finite symmetry transformations are
\begin{equation}
x\rightarrow f(x) ,\hs{3ex}y\rightarrow f'(x)^dy,
\end{equation}
and
\begin{equation}
x\rightarrow x,\hs{3ex}y\rightarrow y+g(x).
\end{equation}

From the infinitesimal transformations \eqref{epsilon1},\eqref{epsilon2},
 we read the generators
\begin{equation}
l_n=-x^{n+1}\partial_x-d(n+1)x^n y\partial_y,
\end{equation}
\begin{equation}
m_n=x^{n+d}\partial_y,
\end{equation}
which satisfy the algebra\eqref{algebra}
  \begin{eqnarray*}
 &&\left[l_n,l_m\right]= (n-m)l_{n+m},\\
&&\left[l_n,m_m\right]= (dn-m)m_{n+m} ,\\
&&\left[m_n,m_m\right] =0.
\end{eqnarray*}
This algebra is analogous to the Witt algebra, and it is called the spin-$d$ Galilean algebra. The central extended one $\tilde{g}=g\oplus C$ has been discussed in section 2.

Now we require that the action of the theory is invariant under the symmetries above
\begin{equation}
\delta S[\delta A_\mu,\delta \bar{A}_\mu]=0
\end{equation}
where
\begin{equation}
\delta A_\mu=\lambda A_\mu,\ \ \ \delta \bar{A}_\mu=d\lambda \bar{A}_\mu+v A_\mu.
\end{equation}
The corresponding currents can be read from
\begin{equation}
\delta S[\delta A_\mu,\delta \bar{A}_\mu]=\int H (J^\mu \delta A_\mu+\bar{J}^\mu\delta \bar{A}_\mu),
\end{equation}
with
\begin{equation}
\bar{J}^\mu A_\mu=0,\ \ \ J^\mu A_\mu+d\bar{J}^\mu \bar{A}_\mu=0.
\end{equation}
In the canonical coordinate $(x,y)$,
\begin{equation}
(\star\bar{J})_x=\bar{h}_x,\ \ (\star\bar{J})_y=\bar{h}_y,\  \ (\star J)_x=h_x,\  \ (\star J)_y=h_y.
\end{equation}
The conditions are simply
\begin{equation}
\bar{h}_y=0,\ \ h_y=-d\bar{h}_x,
\end{equation}
which are exactly the relations \eqref{hbary} and \eqref{hy}.

The other condition is the conservation of the currents\cite{Hofman:2014loa}
\begin{equation}
D_\mu J^\mu_a=0.
\end{equation}
With
\begin{equation}
J^\mu=q^aJ^\mu_a,\ \ \ \bar{J}^\mu=\bar{q}^aJ^\mu_a,
\end{equation}
we have
\begin{equation}
\nabla_\mu J^\mu=0,\ \ \ \nabla_\mu\bar{J}^\mu=0.
\end{equation}
This implies
\begin{equation}
\partial_y \bar{h}_x=0,\ \ \ \partial_y h_x+\partial_x h_y=0.
\end{equation}
This allows us to define  infinitely many conserved charges as in equations \eqref{Mcharges} and \eqref{Qcharges}.

In summary, we have shown that the field theory defined on the Newton-Cartan geometry with anisotropic scaling and boost symmetry indeed possess the conservation currents and charges we need. In the following discussion, we
denote $\bar{h}_x=M(x)$ and $h_x=T(x,y)$.

\section{Quantization}

In this section, we consider how to define the theories on the geometry discussed above. We will use the language in terms of operators  in the discussion, and  we focus on the case with $\ell=d/c$ being integer\footnote{The known examples of warped CFTs and GCA field theories are of this kind. In the case with integer $\ell$, the Cartan algebra is $(L_0,M_0)$. One can discuss the common eigenstates of them and construct the highest weight representation. We can choose $\ell$ to be other values as well, but there is no $M_0$ then. The discussion is similar. From the highest weight representation marked by the eigenvalue of $L_0$, one can also construct the descendant states with $L_{n<0},M_{n<0}$. }. For simplicity, we  set $c=1$ such that $d$ is just an integer.

\subsection{Cylinder Interpretation}
The starting point is the so-called canonical cylinder characterized by a spatial circle $\phi$ and a temporal direction $t$
\begin{equation}
(\phi,t)\sim(\phi+2\pi,t).
\end{equation}
One can get other kinds of spatial circles by tilting $t\rightarrow t+g(x)$. The compactified coordinate is considered in order to eliminate any potential infrared divergence. 
Now, we define the `lightcone coordinates',
\begin{equation}
x=t+\phi,\ \ y=t-\phi.
\end{equation}
We impose the symmetry on the $x,\ y$ directions as discussed before
\begin{equation}
x\rightarrow f(x),\ \ y\rightarrow f'(x)^{d}y
\end{equation}
and
\begin{equation}
x\to x, \ \ y\rightarrow y+g(x),
\end{equation}
with $f(x)$ and $g(x)$ being arbitrary smooth functions of $x$.
Consider the following complex transformation which maps the canonical cylinder to the reference plane
\begin{equation}
z=e^{ix}=e^{t_E-i\phi},\ \ \ \tilde{y}=(iz)^dy,
\end{equation}
where $t_E=-it$ is the Wick-rotated time. We have not considered the tilting of $y$ direction yet. The real time cylinder is capped off at $t=0$ by a reference plane with imaginary time.
\begin{equation}
t_E\rightarrow-\infty,\ \ z\rightarrow 0,
\end{equation}
\begin{equation}
t_E\rightarrow\infty,\ \ z\rightarrow \infty.
\end{equation}
The Hilbert space are defined on the equal imaginary time slices. On the reference plane, this leads to the radial quantization. The `in state' and `out state' are defined by inserting the operators at $t_E=\mp \infty$. On the reference plane, these states are defined at the origin and the radial infinity. One can further put the operators at $y=0$ using the translation symmetry of the $y$ direction.
 The Hamiltonian operator relates different Hilbert space on the canonical cylinder while the dilation on the plane relates the Hilbert spaces on different radial slices (of $x$, but different $y$ ) with each other.

One can inverse the procedure above to get the canonical cylinder from the reference plane. Notice that $z$ provides one real degree of freedom after the continuation, while the other degree of freedom is offered by $y$ instead of the analytical continuation of $\bar{z}$.

The generators of the algebra act on the plane in the following way,
\begin{equation}
L_n=-x^{n+1}\partial_x-d(n+1)x^ny\partial_y,
\end{equation}
\begin{equation}
M_n=x^{n+d}\partial_y.
\end{equation}
The generators
\begin{equation}
L_1,\ L_0,\ L_1,\ M_{-d},\ \cdots\ ,M_d
\end{equation}
can act regularly on each point,  and they generate  the global subgroup.

Now we want to find a set of basis operators filling the representation of the algebra, by the theory of induced representation. The subgroup keeping the  origin invariant is
\begin{equation}
L_0,\ L_{n>0},\ \ M_{-d+1},\ M_{-d+2},\ \cdots.
\end{equation}
The local operators can be labelled by the  eigenvalues $(h_\mO, \xi_\mO)$ of the generators of the Cartan subalgebra $L_0,\ M_0$
\begin{equation}
[L_0,\mO(0,0)]=h_\mO \mO(0,0),\ \ \ [M_0,\mO(0,0)]=\xi_\mO \mO(0,0).
\end{equation}
Requiring $h_\mO$ to be bounded below, one arrives at the highest weight representations
\begin{equation}
[L_n,\mO(0,0)]=0,\ \ [M_n,\mO(0,0)]=0,\ \ \ \mbox{for}\ n>0.
\end{equation}
This defines the primary  operator. One can get the tower of descendant operators by acting $L_{-n},\ M_{-n}$ with $n>0$ on $\mO$.

The operators inserting at the origin give the states,
\begin{equation}
\mO(0,0)|0\rangle\rightarrow |h_\mO,\xi_\mO \rangle.
\end{equation}
This gives a bijection between the states in the Hilbert space at infinitely past and the operators insertion at the origin on the reference plane. Using the commutation relations, the states above fill the representation of the algebra as well. Such representation will be discussed in the following subsection.

The operators at other positions can be got by using the translations,
\begin{equation}
\mO(x,y)=U^{-1}\mO(0,0)U,\ \ \ U=e^{-xL_{-1}+yM_{-d}}.
\end{equation}
In order to compute the commutators $[L_n,\mO(x,y)]$ and $[M_n,\mO(x,y)]$, we notice
 the relations
\bea
[L_n,\mO(x,y)]&=&U^{-1}[UL_nU^{-1},\mO(0,0)]U, \nn\eea
\bea
[M_n,\mO(x,y)]&=&U^{-1}[UM_nU^{-1},\mO(0,0)]U.\nn
\eea
By using the Baker-Campell-Hausdorff (BCH) formula
\begin{equation}
e^{-A}Be^A=B+[B,A]+\frac{1}{2!}[[B,A],A]+\cdots,
\end{equation}
we have
\begin{equation}
UL_nU^{-1}=\sum_{k=0}^{n+1}\frac{(n+1)!}{(n+1-k)!k!}(x^kL_{n-k}-dkyx^{k-1}M_{n+1-d-k}),
\end{equation}
and get
\begin{eqnarray}
[L_n,\mO(x,y)] \nonumber&=&(-x^{n+1}\partial_x-d(n+1)x^ny\partial_y+(n+1)x^n h_\mO+dn(n+1)x^{n-1}\xi_\mO)\mO(x,y) \\
  &+&\sum_{k=n+2-d}^{n}C_{n+1}^k dkyx^{k-1}(M_{n-d-k+1}\mO)(x,y),\ \ \ \mbox{for}\ n\geq -1.
\end{eqnarray}
Similarly, by using
\begin{equation}
UM_nU^{-1}=\sum_{k=0}^{n+d}\frac{(n+d)!}{(n+d-k)!k!}x^k M_{n-k},
\end{equation}
we  find
\begin{equation}
[M_n,\mO(x,y)]=(x^{n+d}\partial_y+C^n_{n+d}x^n\xi_\mO)\mO(x,y)+\sum_{k=n+1}^{n+d-1}C^k_{n+d}x^k (M_{n-k}\mO)(x,y),\ \ \ \mbox{for}\ n\geq -d.
\end{equation}

A special case is $d=0$.  Now $M_0$ does not keep the origin invariant. Nevertheless, $M_0$ is still the generator of the Cartan subalgebra,
\begin{equation}
[M_0,\mO(x,y)]=\xi_\mO \mO(x,y),
\end{equation}
and
\begin{equation}
[M_n,\mO(x,y)]=x^{n}\partial_y\mO(x,y)=x^n\xi_\mO \mO(x,y).
\end{equation}

\subsection{Representation}

The Hilbert space is spanned by the states filling the proper representations of the algebra. The critical assumption is that the spectrum of $L_0$ are bounded below so that we can find the highest weight representations. Starting with an arbitrary state, by acting with the generators $L_n,\ M_n\ (n>0)$, one must reach a state annihilated by all the generators with positive roots. This is the primary state, which is a highest weight state. The Cartan subalgebra is $(L_0,\ M_0)$, so we can find the states with the common eigenstates of $(L_0,\ M_0)$. We consider the case that the primary operators can be diagonalized,
\begin{equation}
L_0|h,\xi\rangle=h|h,\xi\rangle,\ \ \ M_0|h,\xi\rangle=\xi|h,\xi\rangle,
\end{equation}
\begin{equation}
L_n|h,\xi\rangle=0,\ \ \ M_n|h,\xi\rangle=0,\ \ n>0.
\end{equation}
By acting the generators $L_n,\ M_n$ with $n<0$, one gets the descendant states, which are labelled by two vectors $\vec{I},\ \vec{J}$,
\begin{equation}
|\vec{I},\vec{J},h,\xi\rangle=L_{-1}^{I_1}\cdots M_{-1}^{J_1}\cdots|h,\xi\rangle.
\end{equation}
A state is either a primary state or a descendant state, and the Hilbert space is spanned by the modules
\begin{equation}
H=\oplus\sum V_{h,\xi},
\end{equation}
where $V_{h,\xi}$ is the module consisting of a primary state and the tower of all its descendants.  Note that all the null states must be removed.

We have defined the Hilbert space at the origin and discussed the operator-state correspondence. The in-states are
\begin{equation}
|\mO_{in}\rangle=\lim_{z,y\rightarrow0}\mO(z,y)|0\rangle.
\end{equation}
The dilation operator relates one Hilbert space to the others on the reference plane. \\
Now we consider the Hilbert space at the infinity. After the Wick rotation the Hermitian conjugation becomes the reflection of the imaginary time $t_E\rightarrow -t_E$, and on the reference plane
\begin{equation}
z\rightarrow z'=\frac{1}{z^{\ast}},\ \ \ \tilde{y}\rightarrow y'=(\frac{-1}{z^{\ast 2}})^d\tilde{y}\ast,
\end{equation}
where $z^\ast$ is the complex conjugate of $z$.
The dual space is defined by the out-states
\begin{equation}
\langle \mO_{out}|=\lim_{z',y'\rightarrow0}\langle 0|\tilde{\mO}(z',y')
\end{equation}
which can be defined at the infinity on the reference plane, corresponding to the infinite future on the canonical cylinder. \\
The operator $\mO$ transforms as
\begin{equation}
\mO(z',y')=(\frac{-1}{z^{\ast 2}})^h\mO(z,y),
\end{equation}
so the conjugate of the primary operator $\mO$ is
\begin{equation}
\mO^\dagger(z,y)=\mO(\frac{1}{z^{\ast}},(\frac{-1}{z^{\ast 2}})d\tilde{y}\ast)(\frac{-1}{z^{\ast 2}})^h.
\end{equation}
The dual state is,
\begin{equation}
\langle \mO_{out}|=\lim_{z',y'\rightarrow 0}\langle 0|\tilde{\mO}(z',y')=\lim_{z,y\rightarrow0}\langle0|\mO^\dagger(z,y)=|\mO_{in}\rangle^\dagger.
\end{equation}
To map the descendant states to the dual space, consider the mode expansion of the stress tensors,
\bea
M(z)&=&\sum_{n}M_n z^{-n-1-d}, \\
T(z,y)&=&\sum_{n}L_nz^{-n-2}-d\sum_{n}(n+1)yM_{n-d}z^{-n-2}.
\eea
One can impose the condition that $M,\ T$ are Hermitian in the real-time theory, or equivalently  they are real in the imaginary-time theory. This leads to
\begin{equation}
M_n^\dagger=(-1)^{d+1}M_{-n},\hs{3ex}
L_n^\dagger=L_{-n}.
\end{equation}
The inner product and the map are defined by  the adjoint structure above. As
\begin{equation}
L_0^\dagger=L_0,\ \ \ M_0^\dagger=(-1)^{d+1}M_{0},
\end{equation}
$h$ is always real, but $\xi$ is real for odd $d$ and is purely imaginary for even $d$.
Note that the  above conditions are not necessary. In the usual case of CFT$_2$, one imposes such conditions,  with further constraints on the spectrum and central charges, to get a unitary field theory.
In the cases where the theory are not unitary, one can define other adjoint structures.

\section{Two-point Correlation Functions}

In this section, we calculate the correlation functions of the primary operators in the theories with anisotropic scalings.
In the usual CFT, unitarity implies the OPE convergence. For the non-unitary theories we may assume the OPE convergence  to explore potential properties.
In the theories discussed above, considering the radial quantization on the reference plane, it is natural to expect that the operator product expansion is convergent if the theories are unitary. However, such theories cannot be unitary unless  all the $\xi$'s are vanishing. Nevertheless, we assume OPE convergence in such theories.  
With the OPE convergence, the higher point functions can be reduced to lower ones by inserting a complete set of operator basis. Thus the data of such theories are the spectrum and the OPE coefficients.

The correlation functions with imaginary time must be time ordered  in order to be well-defined.  Correspondingly they are radially ordered in the reference plane. We will keep this point in mind without expressing the radial-ordering explicitly.

The vacuum is invariant under the global group discussed in the previous section\footnote{We focus also on the case with integer $\ell$ in this section.}
\begin{equation}
\langle0|G=0,
\end{equation}
where $G$ are the generators of the global subgroup.
Consequently the correlation functions are invariant under the global transformations
\begin{equation}
\langle0|GO(x_1,y_1)O(x_2,y_2)|0\rangle=0
\end{equation}
where
\begin{equation}
G\in\{L_{-1},\ L_0,\ L_1,\ M_{-d},\ \cdots,\ M_d\}.
\end{equation}
Moving $G$ from the left to the right gives the constraints on the two-point functions. For example, the translation symmetries require that the correlation functions must  depend only on $x=x_1-x_2$ and $y=y_1-y_2$.

Let us discuss case by case, setting $c=1$.
The $d=0$ case is special, since the representation is special.  As shown in \cite{Song:2017czq}, there is
\begin{equation}
\langle \mO_1(x,y)\mO_2(0,0)\rangle=d_\mO\delta_{h_1,h_2}\delta_{\xi_1,-\xi_2}\frac{1}{x^{2h_1}}e^{\xi y}.
\end{equation}
For $d=1$, there are no descendant operators involved when doing the local transformations on the primary operators. The two-point function is different from the other cases\cite{Bagchi:2009ca}
\begin{equation}
\langle \mO_1(x,y)\mO_2(0,0)\rangle=d_\mO\delta_{h_1,h_2}\delta_{\xi_1,\xi_2}\frac{1}{x^{2h_1}}e^{2\xi\frac{y}{x}}. \label{2ptb1}
\end{equation}
For $d \geq 2$, the correlation functions become much more involved. The correlation functions of the descendant operators with the primary operators are not vanishing in such cases. Namely we have to consider the following correlation functions
\begin{equation}
f(n,d)=\langle(M_{n}\mO_1)(x,y)\mO_2(0,0)\rangle.
\end{equation}
Solving the constraints  from the invariance of the two-point functions under the global transformations, one gets
\begin{equation}
f(-d+1,d)=-\frac{1}{2}xf(-d,d),
\end{equation}
\begin{equation}
f(n,d)=\frac{(d-1)!(d-n)!}{2(2d-1)!(-n)!}(-1)^{n+d} x^{n+d} f(-d,d),\ \ \ \mbox{for}\ n\in [-d+2,0].
\end{equation}
In the end, one finds
\begin{equation}
\langle \mO_1(x,y)\mO_2(0,0)\rangle=d_\mO\delta_{h_1,h_2}\delta_{\xi_1,(-1)^{d+1}\xi_2}\frac{1}{x^{2h_1}}e^{2C_{2d-1}^d(-1)^{d+1}\xi\frac{y}{x^d}},
\end{equation}
where $C^n_m$ is the binomial coefficient.
When $d=1$, it reduces to the  equation \eqref{2ptb1}.

  \section{Modular Properties}

In this section, we discuss the theories defined on the torus and the modular properties, which means the behaviour of the torus partition funtion under the modular transformation we will discuss below\footnote{We are grateful to Stephane Detournay for suggesting the study of modular properties of the theory. }.
The theories  are defined on the cylinder with the spatial circle
\begin{equation}
(\phi,t)\sim(\phi+2\pi,t).
\end{equation}
Moreover, there is a thermal circle characterizing the temperature and angular potential
\begin{equation}
(\phi,t)\sim(\phi+\alpha,t+\beta).
\end{equation}
The translation charges are
\begin{equation}
Q[\partial_t]=H=M_0^{(cyl)},\ \ \ Q[\partial_\phi]=M=L_0^{(cyl)}.
\end{equation}
They are related to the plane boost and dilation charges
\begin{equation}
x\partial_x+d y\partial_y\rightarrow \partial_\phi,
\end{equation}
\begin{equation}
x\partial_y\rightarrow \partial_t.
\end{equation}
The torus partition function is
\begin{equation}
Z(\alpha,\beta)=\Tr e^{-\beta H-\alpha M}.
\end{equation}
The trace is taken over all states in the Hilbert space of the theory, and $(\alpha,\beta)$ is a pair of the modular parameters characterizing the torus. This is similar with the discussion in \cite{Detournay:2012pc}.

The torus can be described by the fundamental region on the plane
\begin{equation}
(x,y)\sim(x,y)+p(\alpha_1,\alpha_2)+q(\beta_1,\beta_2)
\end{equation}
where $p$ and $q$ are integers. There are different choices of $(\alpha,\beta)$ giving the same lattice and thus the same torus. Assuming that $(\alpha,\beta)$ and $(\gamma,\delta)$ give the same lattice, they are related to each other by an $SL(2,Z)$  transformation
\begin{equation}
\left(
\begin{aligned}
a\ \ &b\\
c\ \ &d
\end{aligned}
\right)\left(
\begin{aligned}
\alpha\\
\beta
\end{aligned}
\right)=
\left(
\begin{aligned}
\gamma\\
\delta
\end{aligned}
\right)
\end{equation}
with
\begin{equation}
ad-bc=1, \hs{3ex}a,b,c,d\in Z.
\end{equation}
And since there is no difference between $(\alpha,\beta)$ and $-(\alpha,\beta)$,   the modular group is actually $SL(2,Z)/Z_2$. Note that the modular group is the isometry group acting on the modular parameters. It describes the different choices of identifications giving the same lattice and thus the same torus. The modular group is independent of the detailed theory. $CFT_2$ also belongs to this case.

The modular group is generated by the $T$ and $S$ transformations. The T-transformation leads to
\begin{equation}
(\alpha,\beta)\rightarrow (\alpha,\beta+\alpha), \hs{3ex}
T=\left( \ba{cc}
1&0\\
1&1
\ea
\right).
\end{equation}
The $S$-transformation leads to
\begin{equation}
(\alpha,\beta)\rightarrow (-\beta,\alpha),\hs{3ex}
S=\left( \ba{cc}
0&-1\\
1&0
\ea
\right).
\end{equation}
Note that the S-transformation exchanges the identifications along the two cycles, instead of the two coordinates.

\subsection{Modular Invariance}

We start with the torus $(\alpha,\beta)$ with two identifications,
\begin{equation}\label{torus1}
(\phi,t)\sim(\phi+2\pi,t)\sim(\phi+\alpha,t+\beta).
\end{equation}
Consider the symmetry transformation of the theory,
\begin{equation}
\phi\rightarrow f(\phi),\ \ \ t\rightarrow f'(\phi)^d t,
\end{equation}
combining with
\begin{equation}
t\rightarrow t+g(\phi),
\end{equation}
where we have set $c=1$ for simplicity.
Under such transformations,
\begin{equation}
(\phi,t)\rightarrow(\phi',t''),\nn
\end{equation}
and
\begin{equation}
(\phi+2\pi,t)\rightarrow (f(\phi+2\pi),f'(\phi+2\pi)^dt+g(\phi+2\pi)).\nn
\end{equation}
\begin{equation}
(\phi+\alpha,t+\beta)\rightarrow (f(\phi+\alpha),f'(\phi+\alpha)^d(t+\beta)+g(\phi+\alpha)).\nn
\end{equation}

We would like to find the symmetry transformations which are consistent with the torus identification.
For arbitrary point $(\phi',t'')$, there should be  two identifications,
\begin{equation}
(f(\phi),t'')\sim(f(\phi+2\pi),f'(\phi+2\pi)^dt+g(\phi+2\pi))\sim(f(\phi+\alpha),f'(\phi+\alpha)^d(t+\beta)+g(\phi+\alpha))
\end{equation}
where
\be
t''=f'(\phi)^dt+g(\phi).
\ee
If the identifications above are proper, $f(\phi+2\pi)-f(\phi)$ and $f(\phi+\alpha)-f(\phi)$ should not depend on $\phi'$, so
\begin{equation}
f(\phi)=\lambda \phi+q
\end{equation}
since $\phi$ is real. The constant shift $q$ of $\phi$ does not matter, and we can set it vanishing and have
\begin{equation}
f(\phi)=\lambda\phi.
\end{equation}
Then the identifications become
\begin{equation}
(\phi',t'')\sim(\phi'+f(2\pi),t''+g(\phi+2\pi)-g(\phi))\sim(\phi'+f(\alpha),t''+\lambda^d\beta+g(\phi+\alpha)-g(\phi)).
\end{equation}
Similarly, $g(\phi+2\pi)-g(\phi)$ should not depend on $\phi'$, so
\begin{equation}
g(\phi)=g_0(\phi)+k\phi+p
\end{equation}
where $g_0(\phi)\sim g_0(\phi+2\pi)$. Moreover, $\lambda^d\beta+g(\phi+\alpha)-g(\phi)$should not depend on $\phi'$, so
\begin{equation}
g_0(\phi)\sim g_0(\phi+\alpha)
\end{equation}
so $g_0$ is a constant and can be absorbed into $p$. The constant shift $p$ does not matter, and we may consider the transformation,
\begin{equation}
g(\phi)=k\phi.
\end{equation}
In short, the transformation functions should be linear, leading to   the identifications
\begin{equation}\label{torus2}
(\phi',t'')\sim(\phi'+f(2\pi),t''+g(2\pi))\sim(\phi'+f(\alpha),t''+\lambda^d\beta+g(\alpha)).
\end{equation}
Moreover, one can do a S-transformation to exchange the two identifications,
\begin{equation}\label{torus3}
(\phi',t'')\sim(\phi'+f(\alpha),t''+\lambda^d\beta+g(\alpha))\sim(\phi'+f(2\pi),t''+g(2\pi))
\end{equation}
This \eqref{torus3} gives the same torus as\eqref{torus2}.
In order to have a well-defined torus identified as in \eqref{torus1} , we should impose the following conditions,
\begin{equation}
f(\alpha)=2\pi,\ \ \ \ \lambda^d\beta+g(\alpha)=0,
\end{equation}
which determine
\begin{equation}
\lambda=\frac{2\pi}{\alpha},\ \ \ \ k=-(\frac{2\pi}{\alpha})^d\frac{\beta}{\alpha}.
\end{equation}
Therefore the allowed symmetry transformations are
\be
f(\phi)=\frac{2\pi}{\alpha} \phi, \hs{3ex} g(\phi)=-(\frac{2\pi}{\alpha})^d\frac{\beta}{\alpha} \phi.
\ee

The new thermal cycle is
\begin{equation}
(\phi,t)\sim(\phi+\alpha',t+\beta')
\end{equation}
where
\begin{equation}
\alpha'=f(2\pi)=-\frac{4\pi^2}{\alpha},\ \ \ \ \beta'=g(2\pi)=-(\frac{2\pi}{\alpha})^{d+1}\beta.
\end{equation}
Note that the transformation of $\phi$ is purely a scaling, not leading to any anomaly in the partition function. When $d\neq 0$, the partition function is invariant under the modular transformation
\begin{equation}
Z(\alpha',\beta')=Z(\alpha,\beta).
\end{equation}
since the modular transformations are re-scaling and Galilean boost, which all belong to the global transformation. However, when $d=0$, the transformation of $t$ may introduce anomaly due to the non-vanishing $c_M$, since the Galilean boost generated by $M_1$ does not belong to the global transformations\cite{Detournay:2012pc}.\\


\subsection{Cardy-like formula: $d\neq 0$}

In this subsection, we calculate the spectrum density at small $\alpha$ or large $\Delta,\xi$ by the saddle point approximation in the cases where $d\neq0$. In the holographic CFT, the degeneracy of the highly excited states could be related to the entropy of the dual configuration. The torus partition function can be recast into the one without vacuum charges
\begin{equation}
\tilde{Z}(\alpha,\beta)=\int_{\Delta_0}^{\infty}\int_{\xi_0}^{\infty}e^{-\alpha\Delta-\beta\xi}\rho(\Delta,\xi)
\end{equation}
where the $\tilde{Z}$ is the partition function defined by
\begin{equation}
\tilde{Z}(\alpha,\beta)=e^{\alpha M_v+\beta H_v}Z(\alpha,\beta).
\end{equation}
Here $M_v$ and $H_v$ are the translation charges of the vacuum. They can be non-vanishing in various theories.
The spectrum density could be read via inverse Laplace transformation
\begin{equation}
\rho(\Delta,\xi)=-\frac{1}{(2\pi)^2}\int d\alpha d\beta e^{\alpha \Delta+\beta\xi} \tilde{Z}(\alpha,\beta).
\end{equation}
Just like in 2D CFT case\cite{Cardy:1986ie}, the key point is to use the modular invariance to reexpress the equation above, and do the saddle point approximation at large $\Delta,\xi$. The modular invariance suggests
\begin{equation}
\tilde{Z}(\alpha,\beta)=e^{(\alpha-\alpha')M_v+(\beta-\beta')H_v}\tilde{Z}(\alpha',\beta').
\end{equation}
For non-vanishing $H_v$ and $M_v$, the saddle point is at
\begin{equation}
\alpha=\frac{2\pi}{(1+\frac{\xi}{H_v})^{\frac{1}{d+1}}},
\end{equation}
and the microcanonical entropy is
\begin{equation}
S(\Delta,\xi)=\log \rho(\Delta,\xi)=\frac{2\pi(\Delta+M_v)}{(\frac{\xi}{H_v}-1)^{\frac{1}{d+1}}}+2\pi(\frac{\xi}{H_v}-1)^{\frac{1}{d+1}}M_v.
\end{equation}
For
\begin{equation}
\xi>>H_v,\ \ \ \ \Delta>>M_v, \nn
\end{equation}
the entropy is
\begin{equation}
S(\Delta,\xi)=2\pi\Delta(\frac{H_v}{\xi})^{\frac{1}{d+1}}+2\pi M_v(\frac{\xi}{H_v})^{\frac{1}{d+1}}. \label{entropy}
\end{equation}
When $d=1,\ M_v=0$, it matches with the result in \cite{Bagchi:2012xr} in which the entropy reproduces the Bekenstein-Hawking entropy of flat cosmological horizon.  For general $d=1$ case, our result match with the one in \cite{Bagchi:2013qva}.

Several remarks are in order:\\
1. We assume that there is no $M$ extension such that the theory is anomaly-free and modular invariant. \\
2.The saddle point approximation is valid. The $\tilde{Z}(\alpha,\beta)$ does vary slowly around the putative saddle point.
Near the saddle point, $\alpha$ is small. The dominant part of the partition function is the contribution from vacuum module. We have already extracted the vacuum charges so that $\tilde{Z}$ approaches a constant as $\alpha\rightarrow 0$. As a result,  the vacuum charges are more important than the central charges, and appear in the entropy formula.\\
3. The relation \eqref{entropy} could be taken as a kind of entropy, or the degeneracy of the states,  if there is no state with negative norm. Otherwise, the integral spectrum density results from the difference between the positive and negative spectrum density. \\
4. In the case $H_v=M_v=0$, the saddle point approximation is not valid since there is no sharp Gaussian contribution around the saddle point.

\subsection{Cardy-like formula: $d=0$}

If the algebra has non-vanishing $M$-extension, the torus partition function is covariant under the modular transformation. There is an anomaly due to the $M$ central charges.
For the infinite dimensional spin-$\ell$-Galilean algebra with finite number of central extensions, the only case with $M$-extension is $d=0$, the warped CFT case. Now there is
\begin{equation}
[M_n,M_m]=c_Mn\delta_{n+m,0}.
\end{equation}
If $g(x)$ is not vanishing, there are anomalies of the currents coming from the $M$-extension\cite{Detournay:2012pc}
\begin{equation}
M(w)=w'^{-1}(M(x)+c_Mg'(x)),
\end{equation}
\begin{equation}
T(w)=w'^{-2}\left(T(x)-\frac{c_T}{12}s(w,x)-g'(x)M(x)-\frac{c_M g'(x)^2}{2}\right),
\end{equation}
where $s(w,x)$ is the Schwarzian derivative
\begin{equation}
s(w,x)=\frac{w'''}{w'}-\frac{3}{2}(\frac{w''}{w'})^2.
\end{equation}
The modular $S$-transformation leads to
\begin{equation}
T\rightarrow \frac{4\pi^2}{\alpha^2}T-\frac{2\pi\beta}{\alpha^2}M+\frac{\beta^2c_M}{2\alpha^2}.
\end{equation}
Therefore, the partition function is covariant under the transformation
\begin{equation}
Z(\alpha,\beta)=e^{\frac{\beta^2c_M}{2\alpha}}Z(\alpha',\beta')
\end{equation}
where
\begin{equation}
\alpha'=-\frac{4\pi^2}{\alpha},\ \ \ \ \beta'=-\frac{2\pi\beta}{\alpha}.
\end{equation}
The partition function without the vacuum charges transforms as
\begin{equation}
\tilde{Z}(\alpha,\beta)=e^{(\alpha-\alpha')M_v+(\beta-\beta')H_v+\frac{\beta^2c_M}{2\alpha}}\tilde{Z}(\alpha',\beta').
\end{equation}
The saddle point is at
\begin{equation}
\beta=-\frac{(\alpha+2\pi)H_v+\alpha\xi}{c_M},\ \ \ \ \alpha=2\pi\frac{\sqrt{H_v^2+2c_MM_v}}{\sqrt{H_v^2-2c_MM_v-2c_M\Delta+2H_v\xi+\xi^2}}.
\end{equation}
The microcanonical ensemble entropy is
\begin{equation}
S(\Delta,\xi)=-2\pi\frac{H_v(H_v+\xi)}{c_M}-\frac{2\pi}{c_M}\sqrt{(H_v^2+2c_MM_v)(-2c_M(M_v+\Delta)+(H_v+\xi)^2)}.
\end{equation}
For
\begin{equation}
\xi>>H_v,\ \ \ \ \Delta>>M_v, \nn
\end{equation}
the entropy is
\begin{equation}
S(\Delta,\xi)=-2\pi\frac{H_v\xi}{c_M}-\frac{2\pi}{c_M}\sqrt{(H_v^2+2c_MM_v)(-2c_M\Delta+\xi^2)},
\end{equation}
which has been obtained in \cite{Detournay:2012pc}.

\section{Conclusion and Discussion}


In the present work we studied a class of general GCFT with anisotropic scaling $x\to \l^c x, y\to \l^d y$.  Under the assumption that the dilation operator is diagonalizable, and has a discrete, non-negative spectrum, we showed in two different ways that the field theories with global translation, Galilean boost and anisotropic scaling could have enhanced symmetries. The first way is to generalize Hofman-Strominger theorem to the case at hand. The global symmetries are enhanced to the infinite dimensional spin-$l$ ($\ell=d/c$) Galilean algebra with possible central extensions. The second way relies on the Newton-Cartan geometry with scaling structure on which the field theory could be defined in a covariant way. Then the enhanced local symmetries could be understood as the consequence that the action of the field theory is invariant under coordinate reparametrization.

Furthermore we discussed the properties of the anisotropic GCFT. We establish the state-operator correspondence by studying the representation of the algebra of the enhanced symmetries. We noticed that when $\ell>1$ the primary operators do not transform covariantly under the local symmetries. And consequently the two-point  correlation functions become more involved, as we had to consider  the correlation functions of a certain set of descendant operators at the same time.

With the Newton-Cartan geometry, we were allowed to consider the theory defined on a torus and discuss the modular properties of the torus partition function. For the theories without $M$-extension, they are modular invariant, while for the theories with $M$-extension they are modular covariant. In both cases, we derived  Cardy-like formulae to count the degeneracy of highly excited states.

Having a covariant formalism, we can go further to explore other properties of such theories. One interesting issue is the scaling anomaly in the partition function. The theory is defined on the equivalent class of the geometry. However, the measure is not invariant under the local transformation\footnote{For the Galilean field theories in higher dimensions, the anomaly issue was studied in \cite{Jensen:2014hqa}. }. An effective action should be given to describe the anomaly of the partition function. In 2D CFT, such effective action is a Liouville action\cite{Polyakov:1981rd}. For the warped CFT, the effective action is a Liouville-type action\cite{Hao2019}. It would be interesting to study the effective action of the anisotropic GCFT. We leave this issue to  the future work. Another interesting problem is to construct explicitly the simple examples realizing the enhance symmetries. This may help us to understand the symmetries better. 

Another important direction is trying to bootstrap these field theories. Some efforts have been made for $l=1$ \cite{Bagchi:2016geg,Bagchi:2017cpu} and $l=1/2$\cite{Goldberger:2014hca}. It will be interesting if one can obtain any dynamical information for these non-relativistic conformal field theories using some well established bootstrap equations with appropriate inputs and reasonable assumptions. One subtle but essential point  is that the theories now are generally non-unitary. To our best knowledge, there are few analytical bootstrap results for non-unitary (conformal) theories. In fact, unitarity is needed for the non-negativity of square of the OPE coefficients, which is crucial for both numerical and analytical bootstrap method. Also, note that the algebras here are generally not semi-simple (while the conformal algebra is), studying the bootstrap in  such theories is nontrivial. Even though the algebra generically include an $SL(2,R)$ sector, the constraints from crossing equation could be different from a simple 1D CFT, which has been studied in SYK model\cite{Qiao:2017xif,Simmons-Duffin:2017nub,Mazac:2018qmi}.

The anisotropic GCFTs  allow us to study the holographic correspondence  beyond AdS/CFT. For the non-relativistic scale invariant field theory, the underlying spacetime is better described by an Newton-Cartan geometry with additional scaling structure. The bulk dual would be at least one dimensional higher. The symmetry consideration may lead to the construction of the bulk dual. For example,
as proposed in \cite{Hofman:2014loa}, the lower spin gravity could be the minimal set of holographic duality of warped CFT. it would be interesting to investigate the holographic dual of a general GCFT with anisotropic scalings.



\vspace*{.5cm}
\noindent {\large{\bf Acknowledgments}} \\

We are grateful to Luis Apolo, Stephane Detournay, Wei Song, Jian-fei Xu for valuable discussions.
The work  was in part supported by NSFC Grant No.~11275010, No.~11335012, No.~11325522 and No. 11735001. We thank  Tsinghua Sanya International Mathematics Forum for hospitality during the workshop \textquotedblleft Black holes and holography\textquotedblright.

\end{document}